\documentclass[preprintnumbers,11pt,a4paper,oneside]{article}
\usepackage{contour}
\usepackage{xcolor}
\usepackage{float}
\usepackage{mathrsfs}
\contourlength{0.05em}

\newcommand{\ttbar}{\ensuremath{t\to bq\bar{q'}~}}

\usepackage{jheppub}
\usepackage{color}
\usepackage{graphicx}
\usepackage{wrapfig,enumerate,slashed}

\usepackage{footmisc}
\usepackage{amsmath}
\usepackage{wasysym} 
\usepackage{graphicx}
\usepackage{subcaption}
\usepackage{color}
\usepackage{comment}
\usepackage{appendix}
\usepackage{slashed}
\usepackage{natbib}

\newcommand{\be}{\begin{equation}}
\newcommand{\ee}{\end{equation}}
\newcommand{\beq}{\begin{equation}}
\newcommand{\eeq}{\end{equation}}

\newcommand{\bee}{\begin{eqnarray}}
\newcommand{\eee}{\end{eqnarray}}
\newcommand{\beeq}{\begin{equation}}
\newcommand{\eeeq}{\end{equation}}

\newcommand{\ket}[1]{|#1 \rangle}

\usepackage{tikz}
\usepackage{orcidlink}

\makeatletter
\gdef\@fpheader{}
\makeatother
\begin{document}

\title{QINNs: Quantum-Informed Neural Networks}

\author[a]{Aritra Bal$^*$\,\orcidlink{0000-0001-6321-5189}\note{Corresponding author.}}
\author[a]{Markus Klute\,\orcidlink{0000-0002-0869-5631}}
\author[b]{Benedikt Maier$^*$\,\orcidlink{0000-0001-5270-7540}}
\author[b,c]{Melik Oughton\,\orcidlink{0009-0002-8869-8521}}
\author[b]{Eric Pezone\,\orcidlink{0009-0000-5882-5898}}
\author[d]{Michael Spannowsky$^*$\,\orcidlink{0000-0002-8362-0576}}

\emailAdd{aritra.bal@kit.edu}
\emailAdd{benedikt.maier@cern.ch}
\emailAdd{michael.spannowsky@durham.ac.uk}

\affiliation[a]{\vspace{0.1cm} Institute of Experimental Particle Physics, Karlsruhe Institute of Technology, 76131 Karlsruhe, Germany}
\affiliation[b]{\vspace{0.1cm} Blackett Laboratory, Imperial College of Science, Technology and Medicine, London, SW7 2AZ, United Kingdom}
\affiliation[c]{\vspace{0.1cm} Department of Physics and Astronomy, University College London, London WC1E 6BT, United Kingdom}
\affiliation[d]{\vspace{0.1cm} Institute for Particle Physics Phenomenology, Durham University, Durham DH1 3LE, United Kingdom}

\preprintA{IPPP/25/60}

\abstract{
Classical deep neural networks can learn rich multi-particle correlations in collider data, but their inductive biases are rarely anchored in physics structure. We propose quantum-informed neural networks (QINNs), a general framework that brings quantum information concepts and quantum observables into purely classical models. While the framework is broad, in this paper, we study one concrete realisation that encodes each particle as a qubit and uses the Quantum Fisher Information Matrix (QFIM) as a compact, basis-independent summary of particle correlations. Using jet tagging as a case study, QFIMs act as lightweight embeddings in graph neural networks, increasing model expressivity and plasticity. The QFIM reveals distinct patterns for QCD and hadronic top jets that align with physical expectations. Thus, QINNs offer a practical, interpretable, and scalable route to quantum-informed analyses, that is, tomography, of particle collisions, particularly by enhancing well-established deep learning approaches.
}

\maketitle

%%%%%%%%%%%%%%%%%%%%%%%%%%%%%%%%%%%%%%%%%%%%%%%%%%%%%%%%%%%%%%
%%%%%%%%%%%%%%%%%%%%%%%%%%%%%%%%%%%%%%%%%%%%%%%%%%%%%%%%%%%%%%
%%%%%%%%%%%%%%%%%%%%%%%%%%%%%%%%%%%%%%%%%%%%%%%%%%%%%%%%%%%%%%

\section{Introduction}
\label{sec:intro}

Jets, collimated cascades of hadrons originating from energetic partons, provide one of the most powerful microscopes with which the LHC explores quantum chromodynamics (QCD) and searches for new physics \cite{Marzani:2019hun}. Their internal substructure encodes a wealth of information about the underlying hard process, colour flow, and hadronisation dynamics~\cite{Gallicchio:2010sw,Larkoski:2013eya,Larkoski:2014gra}. When the initiating parton is a top quark with TeV-scale transverse momentum, its three hadronic decay products overlap so tightly that they are reconstructed as a single top jet~\cite{Butterworth:2008iy,Kaplan:2008ie,Plehn:2009rk}. Disentangling the physics encoded in such highly correlated multi-particle radiation patterns is central to boosted-object tagging and precision top-quark measurement~\cite{Soper:2012pb,Soper:2014rya, Larkoski:2017jix,Kogler:2018hem}. Yet, the complexity of these correlation structures resists simple factorized descriptions and often leaves modern machine-learning classifiers difficult to interpret physically \cite{Cogan:2014oua,Komiske:2018cqr,Qu:2019gqs,Shlomi:2020gdn}.

Jet tomography seeks to overcome this challenge by treating the energy flow inside a jet as a tomographic object, much like medical imaging reconstructs structure from multiple projections~\cite{Dreyer:2018tjj}. We introduce quantum-informed neural networks (QINNs), a general framework that brings quantum information concepts and observables into purely classical models; in this paper, we instantiate it as quantum-informed jet tomography built around the QFIM. The QFIM provides a basis-independent measure of the dynamical sensitivity of a quantum state and thus naturally encodes correlations between jet constituents when events are embedded in a resource-efficient one-particle-one-qubit (1P1Q) representation~\cite{Braunstein:1994zz,Liu:2019xzz}. Importantly, we find that the QFIM extracted from ensembles of QCD and top jets exhibit strikingly different structures: QCD jets appear largely diagonal, reflecting quasi-independent splittings, whereas top jets show pronounced off-diagonal elements characteristic of their three-pronged colour-connected topology. This makes the QFIM an interpretable observable for jet physics. More broadly, we use QINN as an umbrella term for methods that inject quantum structure into classical learning while remaining fully classical at training and inference. A QINN is specified by an encoding of event information into a quantum state, for example, 1P1Q or continuous-variable maps, a choice of quantum observables or metrics, for example, the QFIM, reduced density matrices, entanglement measures, or quantum distances, and an interface to classical models, for example inputs, edge or node embeddings, regularisers, or initialization priors. This work studies one concrete member of that family: a 1P1Q encoding with the QFIM used both as an interpretable observable and as guidance for optimisation.

Recent progress in jet tagging has largely resulted from deep learning methods that learn correlations directly from data. However, these models often remain opaque and can be sensitive to training choices. Alternative approaches, including energy-flow polynomials, correlation functions, and graph-based neural networks, aim to make the underlying structures more transparent. In parallel, the growing field of quantum machine learning has introduced representations such as amplitude and qubit encodings to capture relations in high-dimensional data~\cite{Havlicek:2018nbf,Schuld:2018hrv}. Our study positions itself at the intersection of these developments: by using QFIM as an interpretable, quantum-native probe, we connect quantum geometry with classical machine learning to expose and exploit the correlation patterns that distinguish different jet topologies.  In this context, QFIM also links naturally to optimisation via its role as a quantum metric and in natural-gradient training, and connects to known trainability phenomena in variational circuits \cite{Stokes:2019ads,McClean:2018bwh,Preskill:2018,Blance:2020nhl}. Thus, we first establish the role of QFIM as an untrained classifier, then demonstrate how it can inform and accelerate a simple neural network, and finally extend the framework to GNNs enriched with both QFIM-derived and kinematic features.

This paper makes three contributions. First, it formulates QINN as a general framework for quantum-informed design of classical models. Second, it develops a QFIM-based jet tomography instantiation that exposes class-dependent correlation geometry in an interpretable way. Third, it shows how QFIM-derived quantities act as lightweight graph embeddings and training priors that improve stability and performance in GNNs. Concretely, in Sec~\ref{sec:subjet_qfim} we introduce the QFIM and review the kinematic anatomy of boosted hadronic top decays, set out the qubit map and detail the construction of the QFIM as a correlation matrix.

In Sec.~\ref{sec:QFIM_classifier}, we show that the QFIM can serve directly as a classifier without training. By vectorising the averaged QFIM over event samples, we obtain class prototypes for QCD and top jets. Each new event is then scored according to its proximity to these prototypes, yielding a powerful discriminant even in its untrained form. Building on this, we convert the QFIM-derived feature vector into the input of a shallow neural network with a single trainable encoding layer and linear output. This model learns patterns that closely track the QFIM structure itself, and crucially, initialising the network with the QFIM pattern accelerates convergence and improves optimisation dynamics. In this way, the QFIM does not just separate classes, but also informs the training of neural networks.

Having established the QFIM as a tomographic probe and an informative prior for machine learning, we extend the framework to graphs and GNNs in Sec.~\ref{sec:graphs}. In this representation, jet constituents are nodes and the QFIM entries act as edge features, complemented by kinematic information stored on the nodes. As shown in Sec.~\ref{sec:qgnn}, the resulting quantum-informed GNN captures both the geometry of quantum correlations and the kinematics of the jet, achieving excellent tagging performance and exhibiting faster and smoother training than purely kinematics-based models~\cite{Qu:2019gqs,Shlomi:2020gdn}. Thus, the QFIM acts both as a physically interpretable observable and as a guiding principle for network design.

Finally, we summarise our conclusions in Sec .~\ref {sec:summary}. Our study demonstrates that QINNs for jet tomography unify the interpretability of quantum geometry with the scalability of modern deep learning. By embedding jets into the 1P1Q formalism and exploiting the QFIM, we expose correlation structures that distinguish top jets from the QCD background, accelerate training through informed initialisation, and enhance GNN performance. This study outlines how to integrate quantum observables into collider analyses for classification tasks. It is compatible with near-term quantum hardware and outlines an approach for how to utilise hybrid classical–quantum analysis frameworks.

\section{Subjet-correlations of boosted hadronic jets}
\label{sec:subjet_qfim}

\begin{figure}[h]
    \centering
    \includegraphics[width=0.65\textwidth]{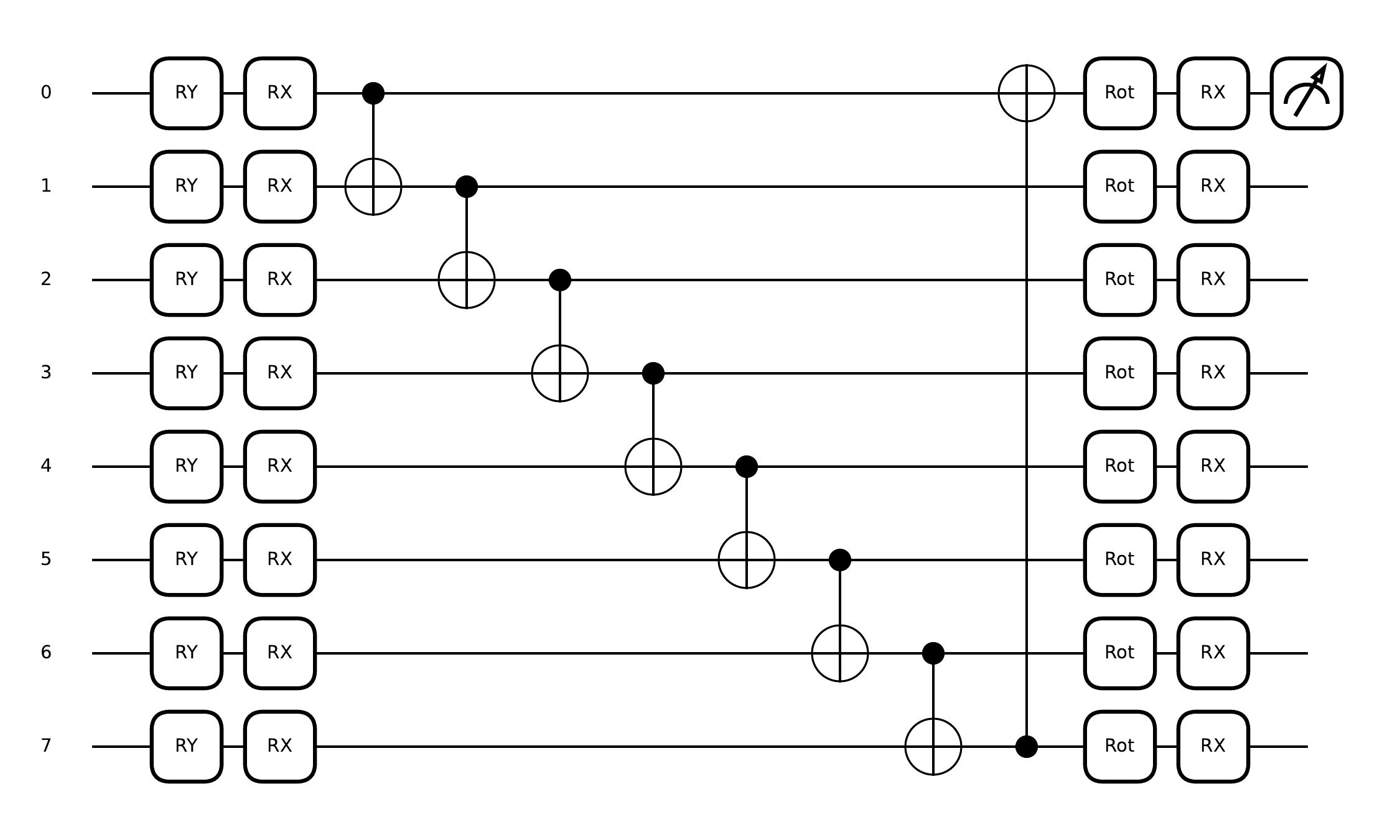}
    \caption{Circuit used to construct the VQC}
    \label{fig:vqc_circuit}
\end{figure}

We want to quantify, in a basis–independent manner, how much of the dynamical information present in a boosted hadronic top jet survives the mapping from classical four–vectors to a quantum state in the 1P1Q scheme.  For this task, we employ the QFIM, a quantity that sits at the intersection of differential geometry and quantum estimation theory \cite{barndorff1998fisher,Braunstein:1994zz}.
On the manifold of pure states endowed with the Fubini–Study metric~\cite{fubini1904,study1905}, the QFIM acts as the local metric tensor; its components therefore measure the squared geodesic distance that the state $|\psi(\boldsymbol\theta)\rangle$ travels under an infinitesimal change of the circuit parameters $\boldsymbol\theta$.  Large entries signal directions in parameter space along which the state moves rapidly, implying high sensitivity and good trainability. In contrast, vanishing entries identify \textit{flat} directions that can give rise to barren plateaus.
Formally, for a variational state $|\psi(\boldsymbol\theta)\rangle$ produced by the circuit shown in Figure~\ref{fig:vqc_circuit}, the QFIM is defined as:
\begin{equation}
    F_{ij}\;=\;4\,\mathrm{Re}\!\Bigl[\,
      \langle\partial_i\psi(\boldsymbol\theta)\,|\,\partial_j\psi(\boldsymbol\theta)\rangle
      \;-\;
      \langle\partial_i\psi(\boldsymbol\theta)\,|\,\psi(\boldsymbol\theta)\rangle\,
      \langle\psi(\boldsymbol\theta)\,|\,\partial_j\psi(\boldsymbol\theta)\rangle
    \Bigr],
    \label{eq:qfim_def}
\end{equation}
where $\partial_i\equiv\partial/\partial\theta_i$.

In our circuit, Figure~\ref{fig:vqc_circuit}, every jet constituent is encoded on its own qubit. Thus, the input space is a tensor product of Hilbert spaces $H_{IN}=\ket{q_1}\otimes\ket{q_2}\dots\otimes\ket{q_N}$, for a system comprising $N$ jet constituents. The trainable parameters 
$\boldsymbol\theta=(\theta_x,\theta_y,\theta_z)$ correspond to single–qubit rotations $R_X(\theta_x)R_Y(\theta_y)R_Z(\theta_z)$ applied after an entangling ring of nearest–neighbor \texttt{CNOT} gates, exactly as described in Ref.~\cite{Bal:2025onequbit}.

Because the final quantum state factorises only without inter–subject correlations, the QFIM automatically picks up physical structures such as colour flow and the three–prong topology of top decays.

A geometric picture helps to relate the QFIM to the LHC observable under study.  Consider two neighbouring parameter points $\boldsymbol\theta$ and $\boldsymbol\theta+\mathrm d\boldsymbol\theta$.  The square of the Fubini–Study line element,
$ \mathrm d s^2 = \tfrac14 \sum_{ij} F_{ij}\,\mathrm d\theta_i\,\mathrm d\theta_j$, quantifies the distinguishability of the corresponding quantum states.  For jets originating from QCD splittings, whose radiation pattern is dominated by one soft/collinear branch, the state typically explores only a lower–dimensional sub-manifold of Hilbert space, leading to a QFIM whose off–diagonal entries are suppressed. In contrast, a boosted hadronic top produces three hard subjets tied together by colour connections. The required interference between three energetic qubits is reflected in sizable off–diagonal elements, especially between parameters that act on qubits representing different decay partons. In other words, the geometry encoded by $F_{ij}$ acts as a "tomographic lens" that magnifies the characteristic triangular colour–flow pattern of top decays.

\begin{figure}[ht]
\centering
\begin{subfigure}{0.48\textwidth}
    \centering
    \includegraphics[width=\textwidth]{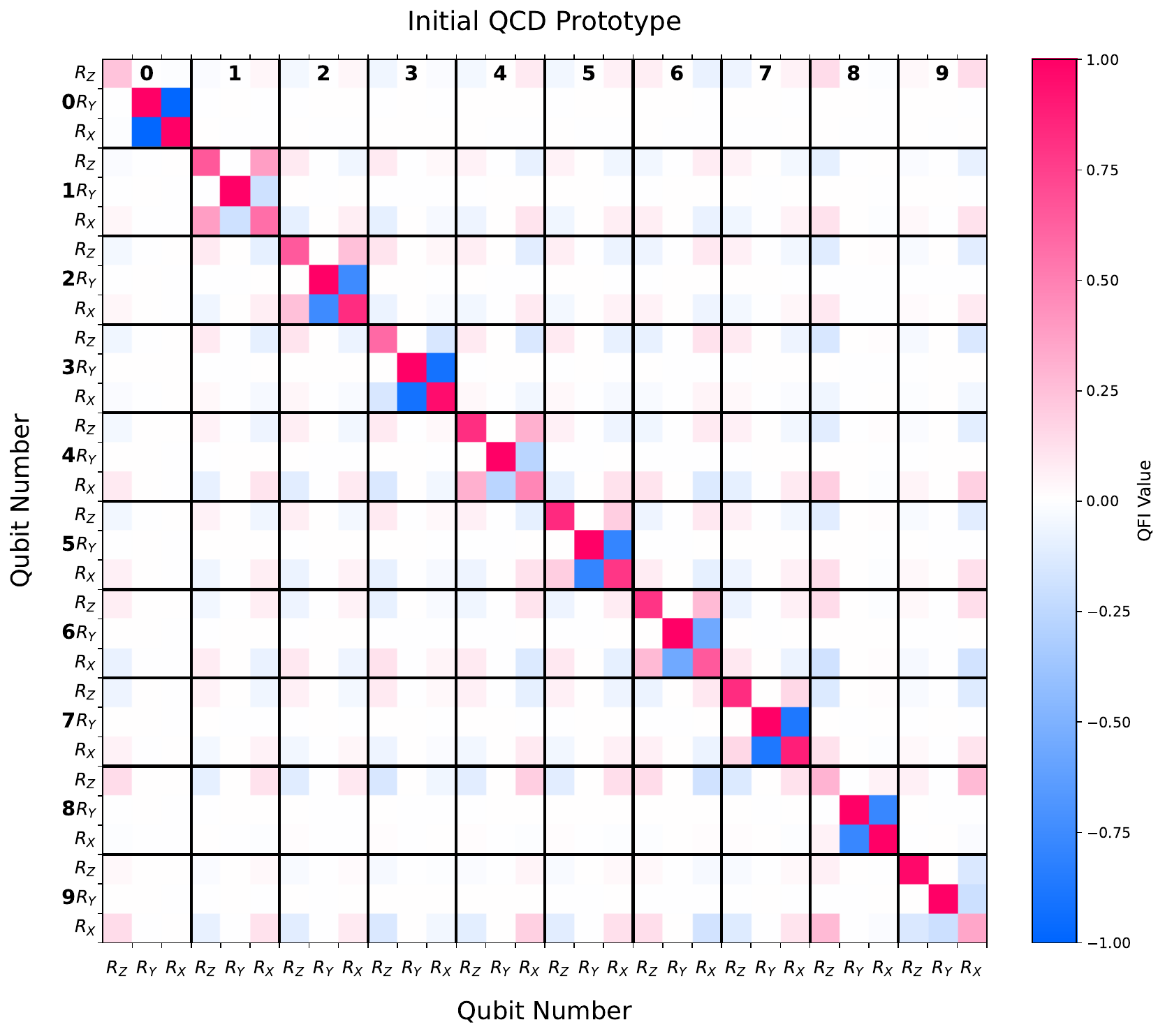}
    \caption{QCD jets}
    \label{fig:qfim_qcd}
\end{subfigure}
\hfill
\begin{subfigure}{0.48\textwidth}
    \centering
    \includegraphics[width=\textwidth]{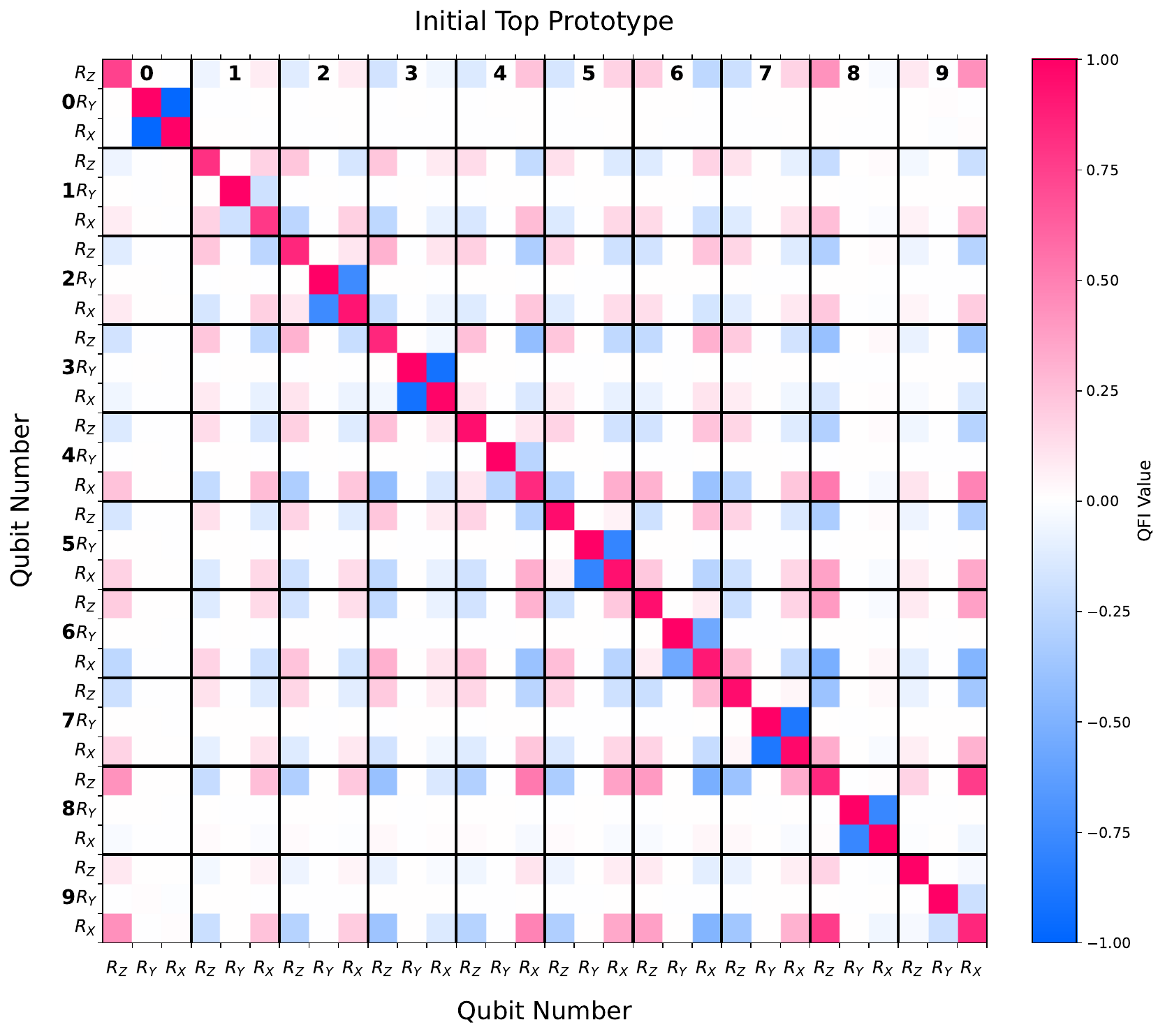}
    \caption{\ttbar decay jets}
    \label{fig:qfim_top}
\end{subfigure}
\caption{QFIM for the 30-parameter Variational Quantum Classifier circuit, implemented in \textsc{PennyLane}~\cite{pennylane}. Left: QFIM averaged over $10^6$ QCD jets. Right: QFIM averaged over $10^6$ jets arising from a top quark jet. The X and Y axes show the parameters corresponding to the trainable rotations applied to each qubit used in the 1P1Q approach, where each jet constituent particle is represented by its own qubit. The matrix elements represent parameter correlations within the quantum circuit, with diagonal elements indicating individual parameter sensitivities and off-diagonal elements capturing inter-parameter dependencies.}
\label{fig:QFI_matrices_QCD_TOP}
\end{figure}

Figure~\ref{fig:QFI_matrices_QCD_TOP} visualises the $30\times30$ QFIM obtained by averaging over ensembles of one million simulated events, separately for ordinary QCD jets (left) and for jets matching a hadronic top hypothesis (right).  For clarity, we omit the classical bias and global scale factor of the VQC, as they do not affect the quantum state itself.  The heat maps corroborate the intuitive picture sketched above: QCD jets populate the diagonal stripes, indicating that each qubit evolves largely independently, whereas top jets exhibit pronounced off–diagonal structure, signalling entanglement between the three leading subjets and the softer, colour–connected radiation that surrounds them.

\begin{figure}[ht]
\centering
\begin{subfigure}{0.48\textwidth}
    \centering
    \includegraphics[width=\textwidth]{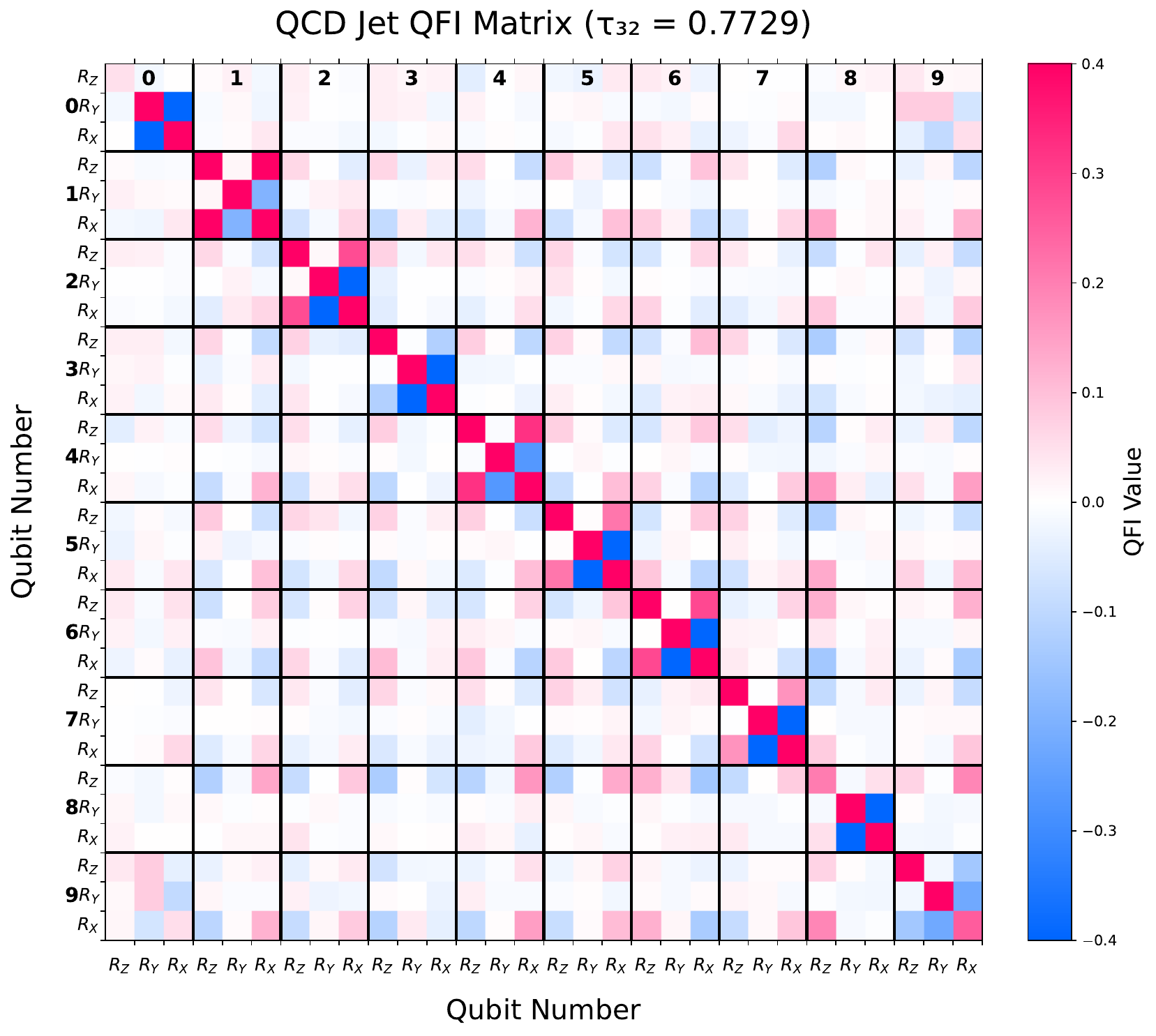}
    \caption{QFIM for a representative QCD jet.}
    \label{fig:qcd_example_qfim}
\end{subfigure}
\hfill
\begin{subfigure}{0.48\textwidth}
    \centering
    \includegraphics[width=\textwidth]{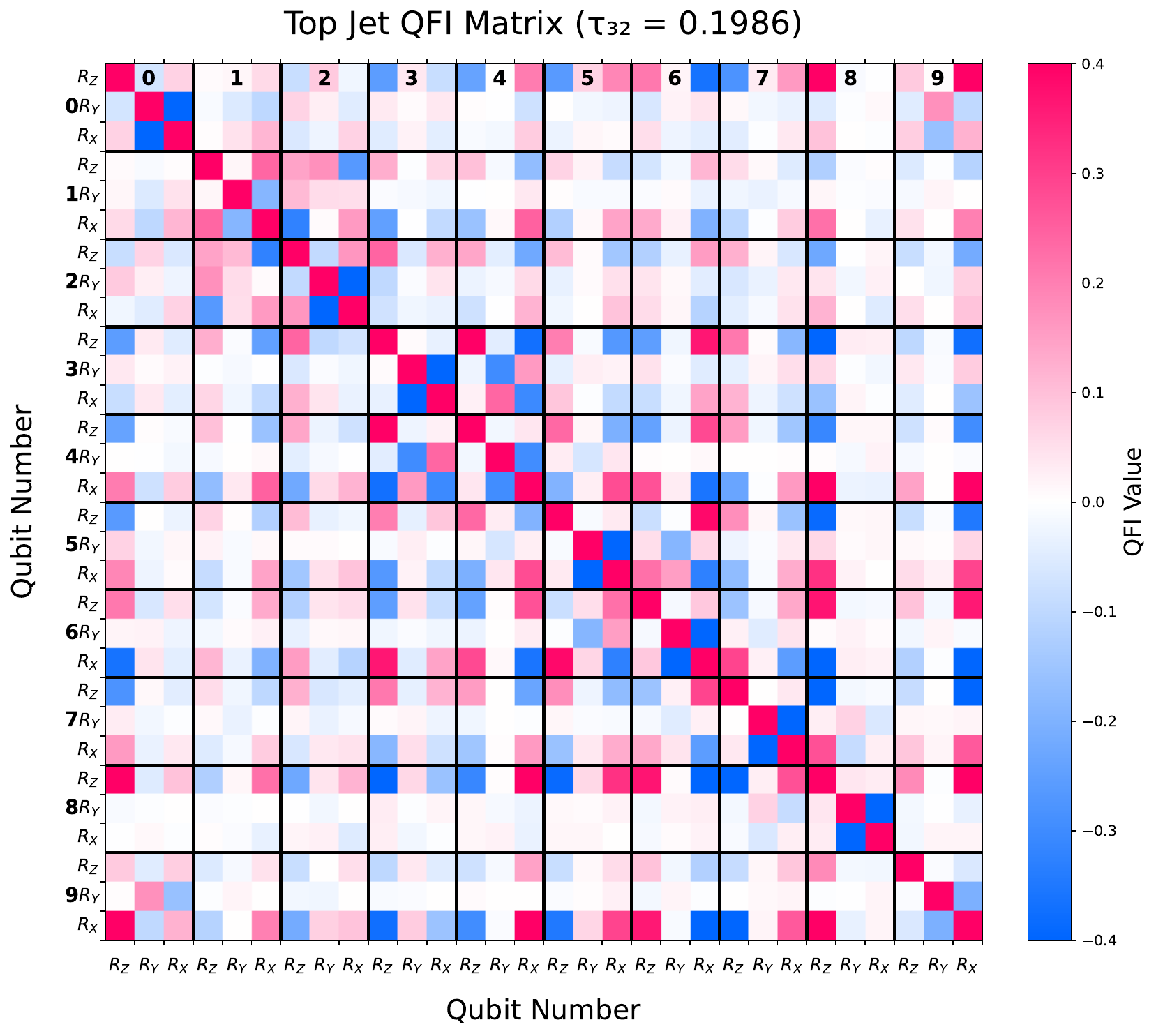}
    \caption{QFIM for a representative top jet.}
    \label{fig:top_example_qfim}
\end{subfigure}

\vspace{0.5cm}

\begin{subfigure}{0.48\textwidth}
    \centering
    \includegraphics[width=\textwidth]{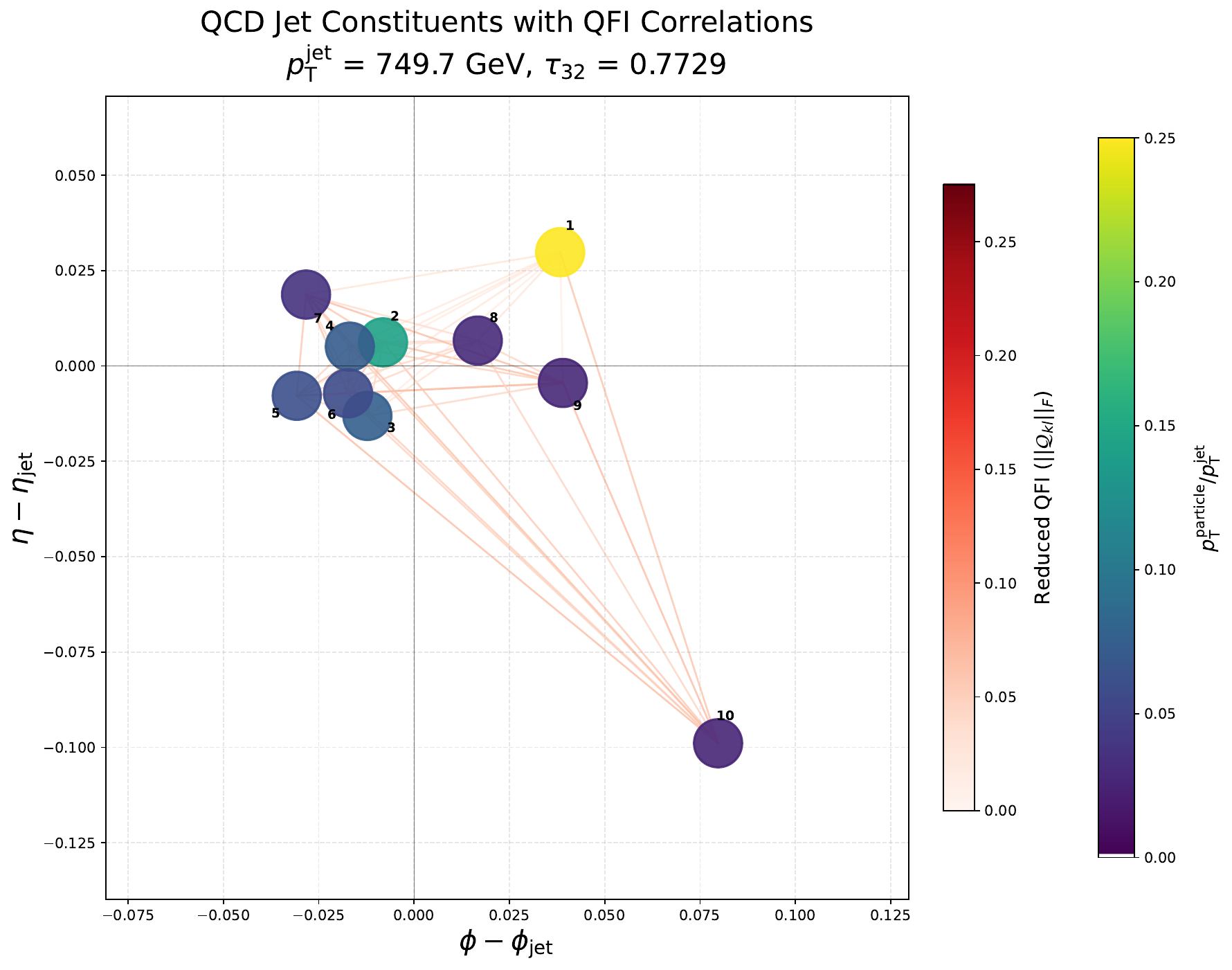}
    \caption{QCD jet constituents in the $\eta$-$\phi$ plane, with the color of the circle representing the fraction of jet $p_\mathrm{T}$ carried by that constituent.}
    \label{fig:qcd_example_constituents}
\end{subfigure}
\hfill
\begin{subfigure}{0.48\textwidth}
    \centering
    \includegraphics[width=\textwidth]{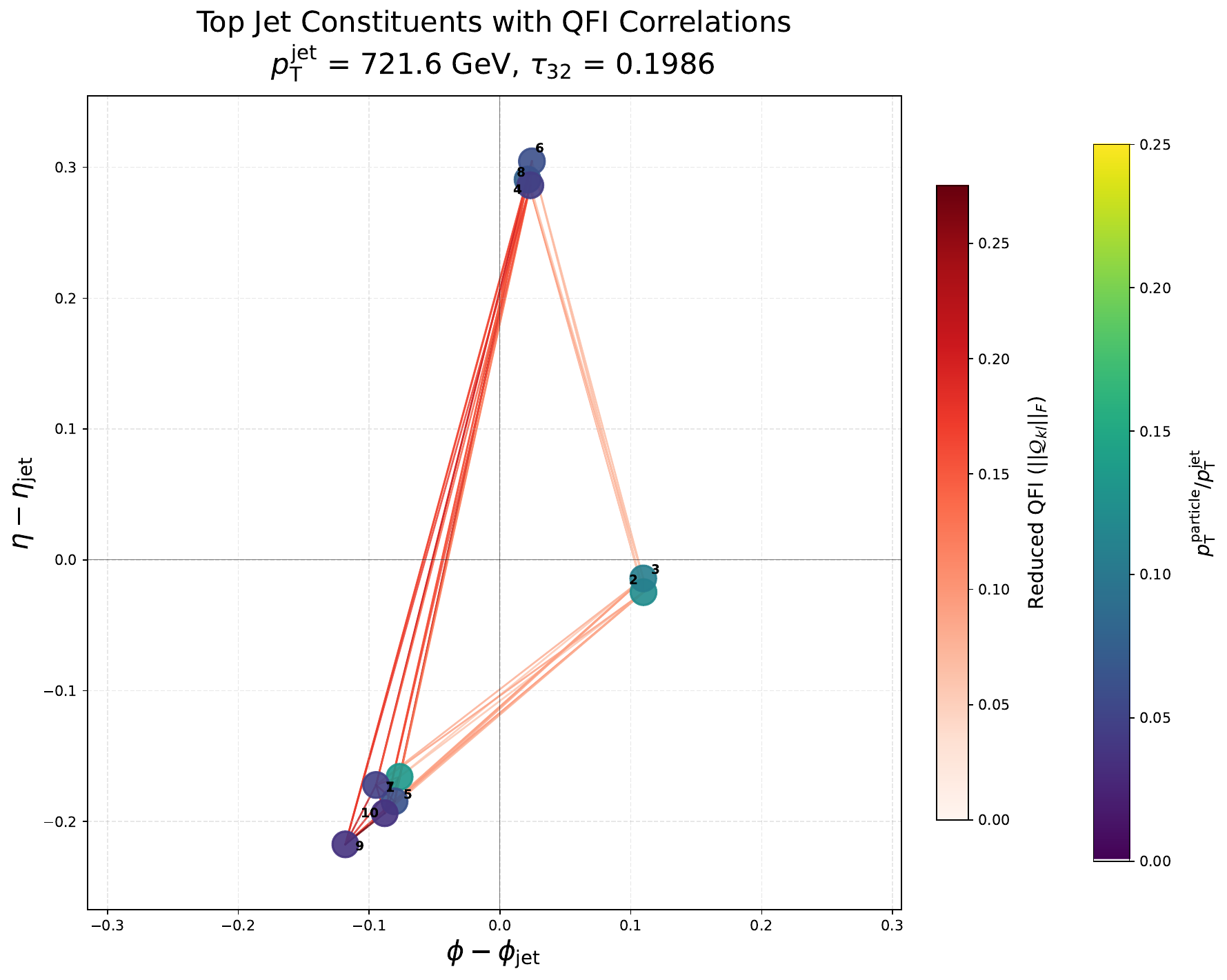}
    \caption{\ttbar jet constituents shown in the $\eta$-$\phi$ plane, with the difference in magnitude apparent when compared to the plot on the left.}
    \label{fig:top_example_constituents}
\end{subfigure}
\caption{QFIMs and spatial constituent distributions for representative QCD (left) and top quark decay (right) jets, chosen using appropriate cuts on the n-subjettiness ratio $\tau_3/\tau_2$~\cite{Thaler:2010tr}, from the $p_\mathrm{T}$ bin $(700,750)\,$GeV. Top row: Full $30 \times 30$ QFIMs with values clipped to the range $[-0.4,0.4]$. Bottom row: Spatial distributions of jet constituents in the $\eta$-$\phi$ plane, with lines connecting constituents representing the strength of correlations derived from the reduced QFIM using the Frobenius norm, and the colour of each circle indicating the fraction of jet $p_\mathrm{T}$ carried by that constituent.}
\label{fig:QFI_LINEPLOT_ALL}
\end{figure}

To facilitate a direct interpretation of inter-particle correlations, we reduce the full $30 \times 30$ QFIM to a $10 \times 10$ representation using the Frobenius norm $||C||_F = \sqrt{\sum_{k,l} C_{kl}^2}$ for $k \in [3i, 3i+3)$ and $l \in [3j, 3j+3)$. In this reduced QFIM, each matrix element $C^\mathrm{reduced}_{kl}$ provides a direct measure of the correlation strength between particles $k$ and $l$, a consequence of the one particle-one qubit encoding scheme. The lines connecting particles in Figures \ref{fig:qcd_example_constituents} and \ref{fig:top_example_constituents} illustrate the intensity of these correlations, revealing that particles belonging to distinct prongs of the top jet exhibit stronger correlations in the QFIM latent space. The corresponding full QFIM for these representative events is shown in Figure \ref{fig:qcd_example_qfim} for the QCD jet and in Figure \ref{fig:top_example_qfim} for the top jet. This analysis demonstrates that the QFIM serves as a powerful diagnostic tool with inherent sensitivity to jet substructure.

From a practical standpoint, computing the full QFIM without block–diagonal approximations is essential.  Only the complete matrix captures the subtle parameter inter-dependencies that differentiate a three–body decay from a quasi–two–body QCD splitting, and only the complete matrix can therefore inform whether training will suffer from barren plateaus or instead benefit from rich gradients.  In subsequent sections, we will exploit precisely these quantum–encoded correlations to inform the training of a GNN that excels at separating boosted top jets from the QCD background of light quark- or gluon-initiated jets.

\section{The QFIM-only Classifier}
\label{sec:QFIM_classifier}
The QFIM, as computed from the Variational Quantum Classifier (VQC) of Ref. \cite{Bal:2025onequbit}, provides an alternative representation of a jet that encodes essentially the same kinematic information contained in the VQC inputs. This observation suggests that the QFIM should, by itself, possess sufficient discriminatory power to distinguish between light quark/gluon-initiated jets and jets originating from the hadronic decay of top quarks. To validate this hypothesis, we begin by constructing a distance-based classifier that leverages the QFIM structure without requiring any training procedure.

The $N \times N$ QFIM, where $N=3\times n_\mathrm{particles}$, is fully symmetric by construction. It therefore contains $D = N(N+1)/2$ independent elements, these being the diagonal entries and either the upper or lower triangular components. Thus, each event with $N$ particles yields a $3N\times3N$ QFIM $F$ arranged as $N\times N$ blocks $F^{(ij)}\in\mathbb{R}^{3\times3}$ that quantify the joint sensitivity of particles $i$ and $j$, see Figs.~\ref{fig:qcd_example_qfim} and \ref{fig:top_example_qfim}. For visualisation, we compress each block to a single edge weight $C_{ij}=\lVert F^{(ij)}\rVert_F$ and draw the constituents in the $\eta,\phi$ plane with an edge whose colour intensity and thickness scale with $C_{ij}$, while node colour reflects momentum fraction, Figs.\ref{fig:qcd_example_constituents} and \ref{fig:top_example_constituents}. Bright off-diagonal structure in the matrix corresponds to strong cross-prong edges in the graph. QCD jets are dominated by near diagonal entries, Fig.~\ref{fig:qcd_example_qfim}, indicating correlations confined within one prong and short-range links among collinear emissions; hadronic top jets exhibit pronounced off-diagonal blocks that tie the $W\to q\bar q^\prime$ pair and the $b$ to the $W$ system, producing a triangular network of edges across the three subjets, Fig.~\ref{fig:top_example_qfim}. In this way, the QFIM provides an interpretable bridge between quantum geometry and jet kinematics, and the reduced $C_{ij}$ graph visualises the correlation patterns that guide message passing in our GNNs of Secs.~\ref{sec:graphs} and \ref{sec:qgnn}.

In this $D$-dimensional feature space, we compute two mean representative vectors, one for each jet class, which we designate as the class prototypes. For each sample in the dataset, we then calculate the squared $L_2$ distances to these class prototypes, denoted as $d_{\mathrm{QCD}}$ and $d_{\mathrm{top}}$, and define a score metric as $S = d_{\mathrm{QCD}} - d_{\mathrm{top}}$. This score is evidently large and positive when the jet in question lies in close proximity to the top jet prototype in feature space, and conversely becomes large and negative for jets similar to the QCD prototype. As such, $S$ naturally serves as a classification score that can be directly converted to a probability via the sigmoid function, enabling the calculation of standard performance metrics such as accuracy and the construction of receiver operating characteristic (ROC) curves.

\begin{figure}[htbp]
    \centering
    \begin{subfigure}{0.48\textwidth}
        \centering
        \includegraphics[width=\textwidth]{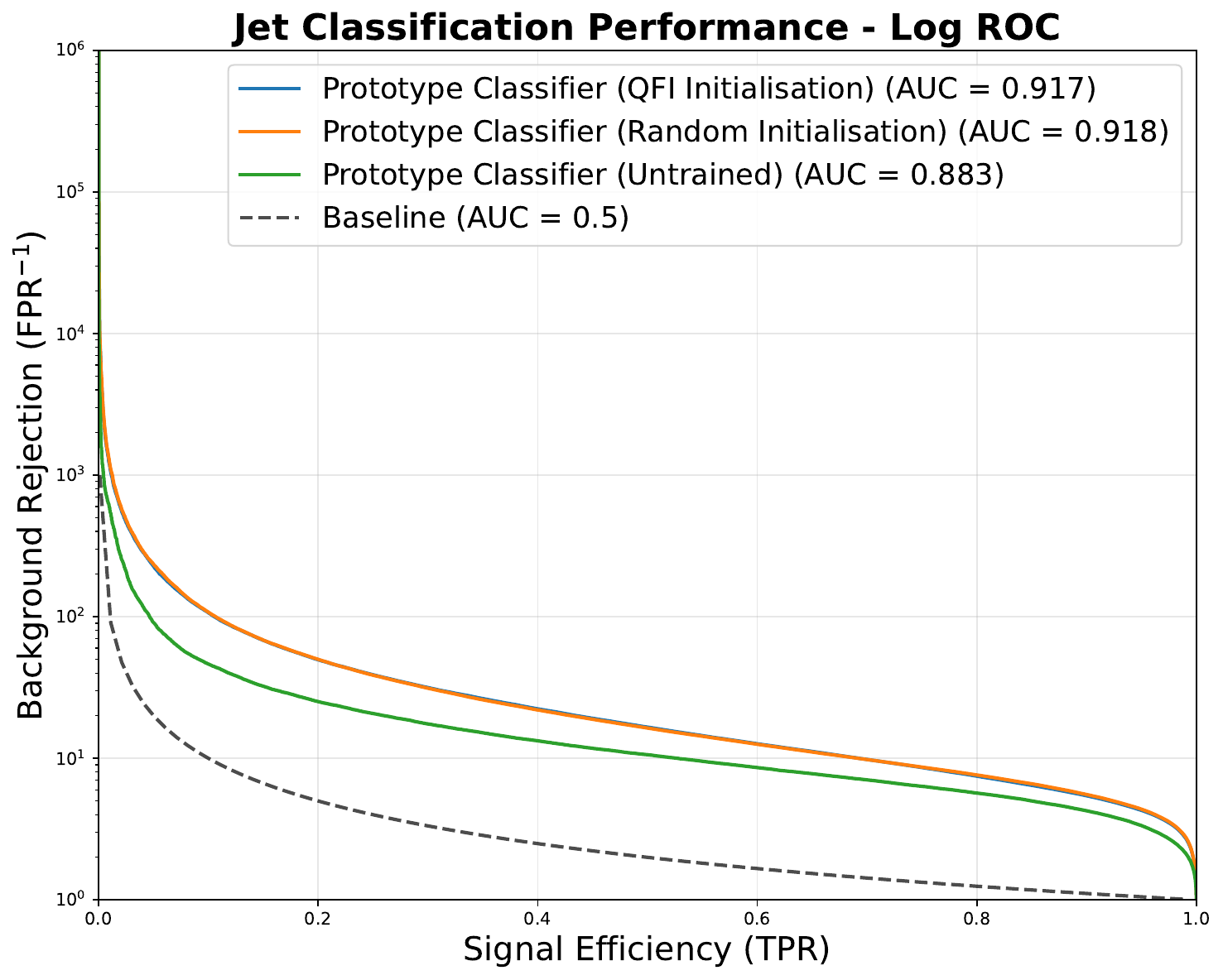}
        \caption{ROC curves comparing the performance of QFIM-initialised and randomly initialised neural network classifiers, both achieving a roughly similar level of performance that is better than the untrained baseline.}
        \label{fig:ROC_QFI_only_NN}    
    \end{subfigure}
    \hfill
    \begin{subfigure}{0.48\textwidth}
        \centering
        \includegraphics[width=\textwidth]{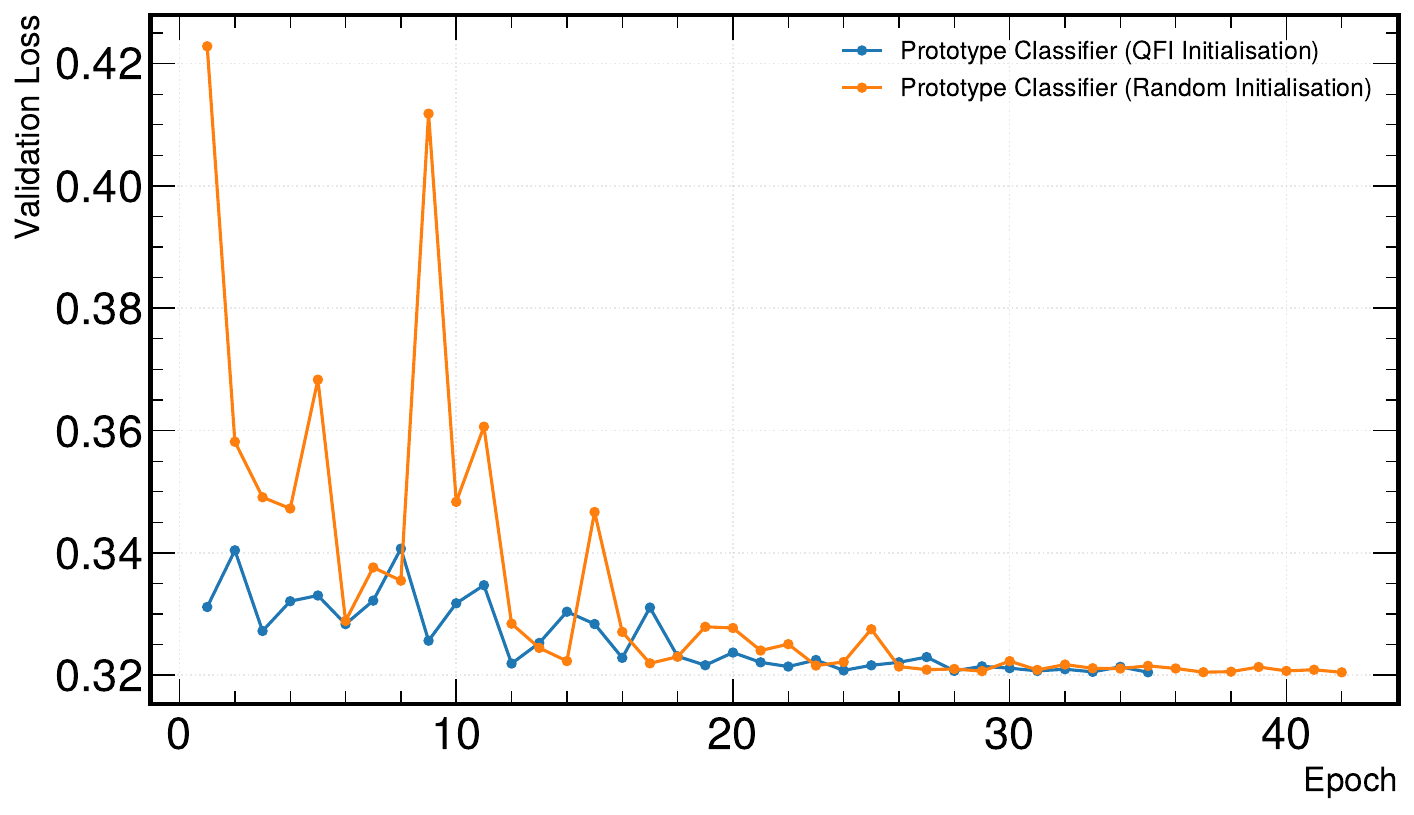}
        \caption{Loss evolution during training for QFIM-initialised (stable, rapid convergence) versus randomly initialised (unstable, slow convergence) neural network classifiers.}
        \label{fig:loss_evolution_QFI_only_NN}
    \end{subfigure}
    
    \caption{Performance comparison of QFIM-initialised versus randomly initialised NN-based classifiers, with the untrained classifier serving as a baseline.}
    \label{fig:QFI_only_NN_performance}
\end{figure}

This naive classifier achieves an area under the curve (AUC) score of 0.883 without any training, as shown in Figure \ref{fig:ROC_QFI_only_NN}. This performance is nearly identical to that of the VQC, thereby validating our assertion that the QFIM effectively encodes the learning outcome of the aforementioned VQC.

Having established the efficacy of the QFIM as a classifier representation, we proceed to demonstrate that the QFIM is already close to being the ideal classifier in our $D$-dimensional feature space, which constitutes a fully quantum representation of the jet, and that utilising it as an initialisation point leads to substantially faster convergence and more stable loss evolution during training. To this end, we employ a shallow single-layer neural network conditioned on the individual jet QFIMs to learn the optimal class prototypes that best separate the two jet classes. Classification is performed using a distance-based approach analogous to that described above. The objective function is chosen to be the margin ranking loss~\cite{pytorch-marginrankingloss} with a margin value of 1.0, with targets appropriately calculated from the truth labels (0 for QCD jets and 1 for top jets).

We consider two initialisation strategies for the class prototypes. In the first instance, the prototypes are initialised to the mean QFIMs for the two classes, the same values used in the naive classifier. For the baseline comparison, we initialise the class prototypes by randomly sampling from a uniform probability distribution in the interval $(-1, 1)$, thereby maintaining initial parity with the range of the QFIM, which is bounded within $(-1, 1)$. In both cases, however, the learned class prototypes are allowed to be unbounded during training, enabling the network to explore the full parameter space.

The results for these two initialisation strategies are presented in Figure \ref{fig:QFI_only_NN_performance}. As shown in Figure~\ref{fig:ROC_QFI_only_NN}, both networks attain relatively similar classification performance, achieving an AUC score of 0.917, a modest but noticeable improvement over the untrained case described previously. However, the convergence behaviour differs starkly between the two approaches, as illustrated in Figure~\ref{fig:loss_evolution_QFI_only_NN}. The QFIM-initialised network reaches loss values close to the final converged value after just one epoch and exhibits stable loss evolution throughout training. In contrast, the randomly initialised network requires more than 20 epochs to attain loss values in the same region, and moreover, experiences unstable loss dynamics during this initial training regime. Only after approximately 30 epochs does the loss stabilise and begin slowly converging toward the level achieved by the QFIM-initialised classifier. The network is updated for each batch of 256 samples. For a training dataset of size $10^6$, the network parameters are updated approximately $4000$ times per epoch. 

\begin{figure}[htbp]
    \centering
    \begin{subfigure}{0.48\textwidth}
        \centering
        \includegraphics[width=\textwidth]{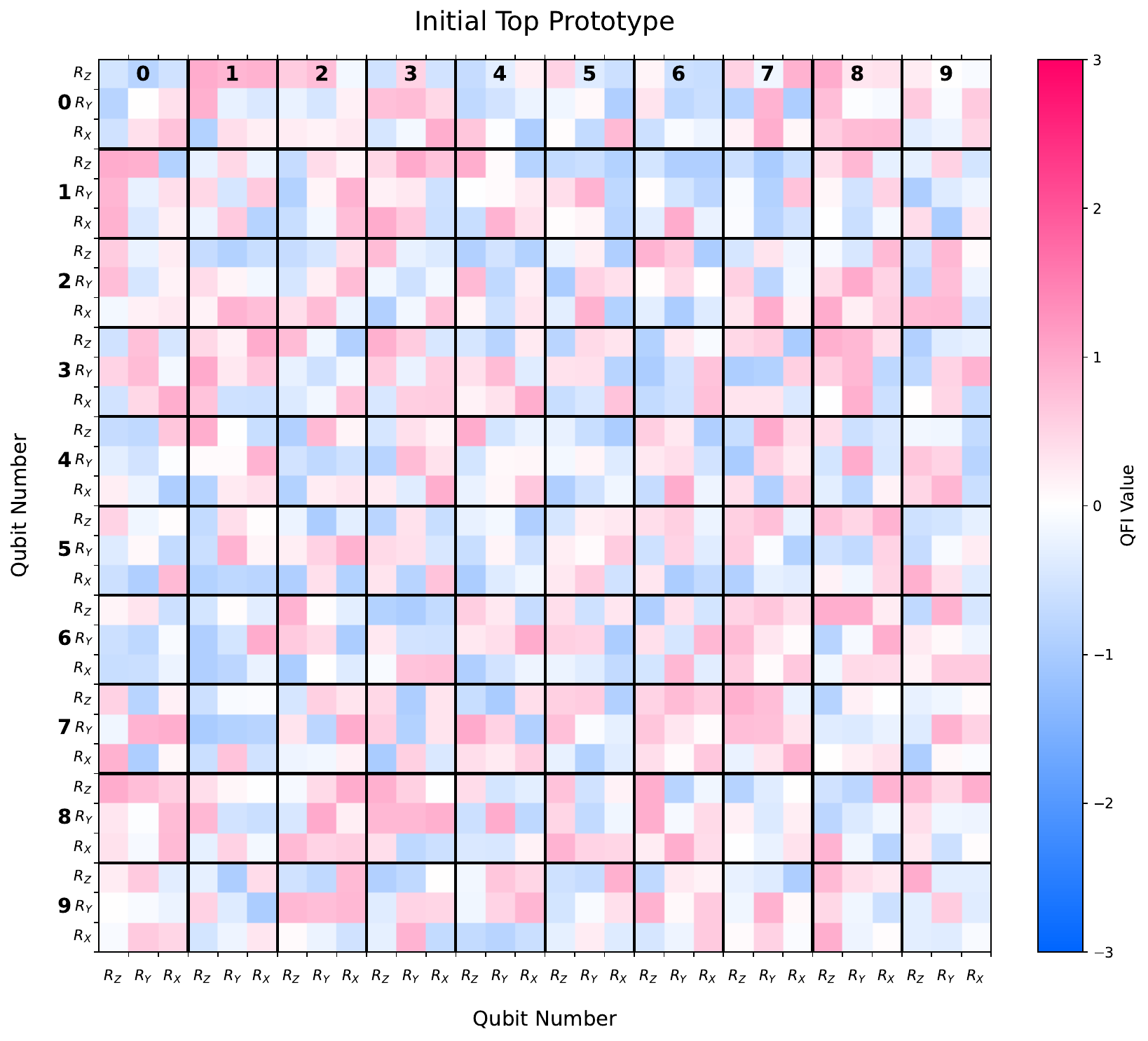}
        \caption{Randomly initialised \ttbar class prototype, drawn from the uniform probability distribution in $[0,1].$}
        \label{fig:random_ttbar_initialisation}    
    \end{subfigure}
    \hfill
    \begin{subfigure}{0.48\textwidth}
        \centering
        \includegraphics[width=\textwidth]{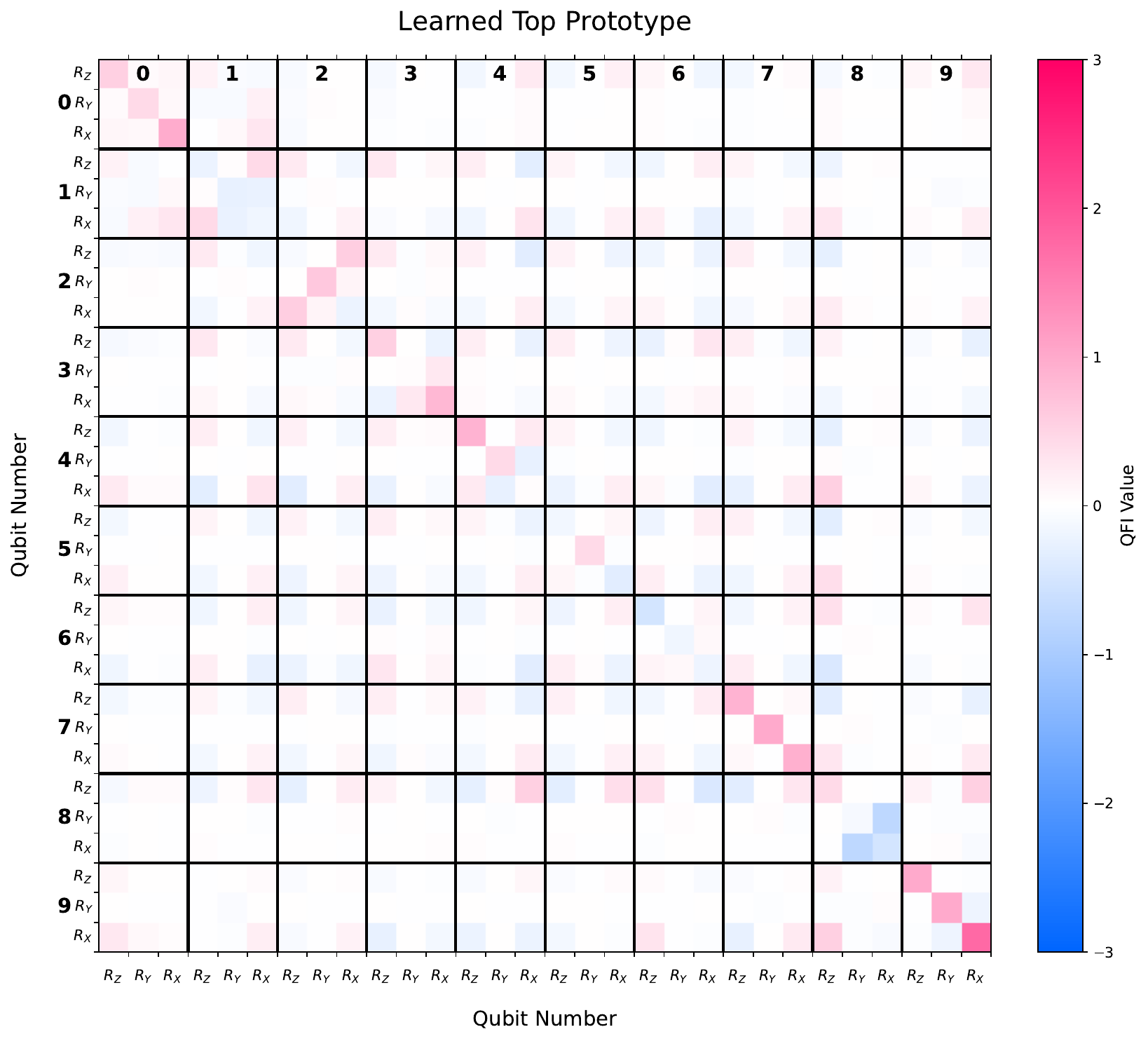}
        \caption{Final learned \ttbar class prototype, with the values allowed to be unbounded, so as to explore the entire possible phase space.}
        \label{fig:final_ttbar_prototype}
    \end{subfigure}
    
    \caption{Comparison of the randomly initialised and learned final class prototypes for \ttbar jets, demonstrating convergence to a structure similar to the mean QFIM. The range of the colour bars is increased so as to account for the unbounded range.}
    \label{fig:QFI_evolution}
\end{figure}

To further illustrate this point, we visualise the randomly initialised class prototype and the final learned class prototype in Figure~\ref{fig:QFI_evolution} for the case of top jets. Remarkably, despite the random initialisation, the optimal class prototypes that best separate the two classes are very similar in structure to the mean QFIMs computed initially. This observation shows that the QFIM is indeed close to being the ideal separator between the two jet classes in this feature space.

The trainable neural networks described above are implemented in PyTorch. Optimisation is performed using the \textsc{Adam} optimiser with an initial learning rate of 0.01, which is allowed to decay by a factor of 2 for every five epochs during which no improvement in the validation AUC is observed (defined as an improvement of less than 0.001). Early stopping is triggered when no such improvement in the validation AUC occurs over a period of 30 consecutive epochs. For all experiments, the training dataset comprises 1 million jets, the validation dataset (applicable only to the trainable case) consists of 200,000 jets, and the test dataset contains 1.2 million jets. In all cases, the datasets are equally divided between the two jet classes to ensure balanced training and evaluation.

\section{Exploiting the constituent particle correlations with graphs}
\label{sec:graphs}

A GNN~\cite{scarselli2008gnn} is a neural network architecture that performs computation directly on graph structures, operating on vertices and edges instead of regular grids like images or sequences.  Each vertex $v_i$ carries a feature vector $h_i^{(0)}$ (here the kinematic attributes of a single subjet, for example its rapidity, azimuth and transverse momentum). In contrast, each edge $(i,j)$ is endowed with a weight or feature vector $e_{ij}$ that influences how strongly the two vertices interact.  The core operation of the network is message passing: at layer $l$ every vertex gathers \textit{messages} sent along the incident edges, aggregates them, and combines the result with its own state to produce an updated feature vector $h_i^{(l+1)}$.  In algebraic shorthand, a broad class of GNNs can be described by the fundamental graph message-passing equation:
\begin{equation}
h_i^{(l+1)} \;=\; \phi^{(l)}\left(h_i^{(l)},\,\underset{j\in\mathcal N(i)}{\textsc{Agg}}\;\psi^{(l)}\!\bigl(h_i^{(l)},\,h_j^{(l)},\,e_{ij}\bigr)\right),
\label{eq:graph_message}
\end{equation}
where $\psi^{(l)}$ is a learnable function that builds a message from the states of the two endpoints (these being the vertices themselves) and their connecting edge. \textsc{Agg} is a permutation‐invariant reduction such as a sum or mean, and $\phi^{(l)}$ is a learnable nonlinear update.  Stacking several such layers lets information flow non-locally across the graph, and a final readout function, which in practice is also a relatively shallow neural network with a single output, maps the set of vertex states to an event-level score. In this case, this is the probability that the jet arose from a hadronic top decay rather than the QCD background.

The entries $C_{kl}$ of the QFIM provide a natural choice for the edge feature vectors $e_{ij}$ in the graph message function of Eq.~\ref{eq:graph_message}. This direct correspondence, in the form of a sub-matrix of the full QFIM, leverages its role in quantifying parameter estimation sensitivity to establish meaningful connectivity patterns in the GNN architecture. A large value indicates that two constituents are kinematically and dynamically linked, and vice versa.  By interpreting every jet constituent particle as a vertex and every non-zero $C_{kl}$ as the feature vector of an edge between $v_i$ and $v_j$, the QFIM can play a role similar to the adjacency matrix $\mathbf A$ of the graph, optionally normalised or thresholded to control sparsity. Since the circuit of the VQC first introduced in Ref. \cite{Bal:2025onequbit} uses three trainable parameters per qubit, whereas $C_{kl} ~\forall ~ k~\in~ [3i,3i+3),~l ~\in ~[3j,3j+3)$ represents a $3\times3$ sub-matrix of the QFIM, corresponding to the two qubits $i$ and $j$. During the first message-passing step, the vector quantity $\mathbf{e}=C_{ij}$ modulates how strongly vertex $j$’s features affect vertex $i$’s update; a large positive value allows a substantial flow of information, while a small value effectively decouples the pair.  

Once this graph representation is fixed, training proceeds exactly like other supervised classification tasks. Simulated or labelled real events provide examples of top-quark jets and QCD jets. For each event, the QFIM controls the graph’s geometry and the interactions between nodes, the constituent kinematics furnish the initial node features, and the network parameters are adjusted to minimise a loss such as the cross-entropy between the predicted label and the truth. During inference, the process is repeated without gradient updates: construct the matrix for the unseen jet, feed the graph through the trained GNN, and obtain a probability that the event is top-like. Because the correlation structure of a boosted hadronic top, containing three hard subjets with characteristic angular separations and energy sharing, differs systematically from that of a QCD jet, the message-passing layers learn discriminative patterns that capture precisely the physics encoded in the QFIM. In this way, the visually compelling heat-map we began with is elevated from an exploratory diagnostic to the central mathematical object that drives a robust, fully differentiable tagging algorithm.

\subsection{Tomography-enhanced graph classifiers for top-tagging}

The QFIM effectively captures the correlations between jet constituent particles in a manner that enhances both the performance and convergence properties of GNNs. Remarkably, this beneficial effect can be observed without the computational overhead of actually training the network and updating its parameters. Instead, the QFIM can be directly incorporated into the information stored at graph vertices and utilised in the graph message passing through appropriately defined mathematical transformations of the input variables, specifically the jet constituent kinematics.

This approach enables the construction of a powerful yet untrained graph-based architecture that operates solely on the jet constituent particle kinematics and the QFIM to perform inference using an appropriate distance metric such as the Mahalanobis distance~\cite{Mahalanobis1936}. The novelty of the approach proposed here lies in the inclusion of the QFIM during the graph construction phase, which, as demonstrated in the following section, substantially enhances the discriminative power between jets originating from hadronic top quark decays and the overwhelming QCD background. This improvement is quantified through a noticeable increase in standard classification metrics, such as the AUC score and overall classification accuracy.

We adopt the input features established in the 1P1Q approach, utilising the kinematic variables associated with each jet constituent. We utilise jet samples from the \textsc{JetClass} dataset~\cite{JetClass}, which contains up to 100 million jets, divided into ten classes. For the purpose of this study, we require only the jets initiated by a light quark or gluon class, or from a \ttbar decay. Each jet in turn is represented by up to the ten highest-$p_{\mathrm{T}}$ constituents, thereby defining each jet as a graph containing at most 10 vertices. Since all vertices are nominally connected through directed edges, this configuration yields a maximum of $10^2 = 100$ edges per graph. The edge features, which determine the extent of interaction between vertices along these connections, are governed by the QFIM as mentioned in the previous paragraph. The input features for each constituent are therefore defined as relative kinematic variables with respect to the parent jet:

\begin{align}
p_{\mathrm{T},i}^{(\mathrm{rel})} &= \frac{p_{\mathrm{T},i}}{p_{\mathrm{T},\text{jet}}} \nonumber\\
\eta_i^{(\mathrm{rel})} &= \eta_i - \eta_{\text{jet}} \nonumber\\
\phi_i^{(\mathrm{rel})} &= \phi_i - \phi_{\text{jet}}
\label{eq:input_feature_definition}
\end{align}

The initial state of the vertices comprising the graph is not directly defined by the aforementioned input variables, but rather by the following mathematical transformations of these variables, specifically designed to maximise the separation between the two classes:

\begin{equation}
\begin{pmatrix}
v_1 \\
v_2 \\
v_3 \\
v_4
\end{pmatrix}
=
\begin{pmatrix}
\frac{\Delta p_{\mathrm{T},ij}}{p_{\mathrm{T},i}^{(\mathrm{rel})} + p_{\mathrm{T},j}^{(\mathrm{rel})} + \epsilon} \\[0.3em]
\frac{\Delta \eta_{ij}}{(1 + \Delta R_{ij})^3} \\[0.3em]
\frac{\Delta \phi_{ij}}{(1 + \Delta R_{ij})^3} \\[0.3em]
\Delta R_{ij}
\end{pmatrix}
\label{eq:vertex_features_baseline}
\end{equation}

The $\Delta p_{\mathrm{T}}$, $\Delta \eta$, and $\Delta \phi$ terms are calculated using the relative kinematic variables defined in Eq.~\ref{eq:input_feature_definition}. However, the nature of these transformations ensures that the specific choice of reference frame does not affect the discriminative power. The $\epsilon$ term, chosen to have a small value of $10^{-4}$, in the denominator of $v_1$ ensures stability and prevents divergences during inference. Lastly, $\Delta R_{ij} = \sqrt{(\Delta \eta_{ij})^2 + (\Delta \phi_{ij})^2}$ represents the standard distance metric between two constituents in the $\eta$-$\phi$ plane.

A graph utilising these four features alone serves as our non-enhanced baseline classifier. Our approach proposes to increase the vertex feature dimensionality from 4 to 7 by incorporating additional features constructed through quantum information enhancement:

\begin{equation}
\begin{pmatrix}
v_5 \\
v_6 \\
v_7
\end{pmatrix}
=
\mathbf{e} \cdot
\begin{pmatrix}
v_1 \\
v_2 \\
v_3
\end{pmatrix}
\label{eq:qfi_augmentation}
\end{equation}

Here, $\mathbf{e}$ is a $3 \times 3 = C_{kl}$ is the sub-matrix of the complete $30 \times 30$ QFIM (for a jet represented using 10 particles), as described in the previous section. It accurately captures the quantum correlations between the particles at vertices $i$ and $j$ that are connected by the edge $e_{ij}$. This approach, therefore, constitutes our quantum information-enhanced graph classifier, which will be demonstrated to exhibit clear superiority over the classical benchmark.

The function $\psi^{(l)}$ of Eq.~\ref{eq:graph_message} can therefore be viewed as a mathematical transform that converts the vertex features defined in Equation \ref{eq:input_feature_definition} and the QFIM sub-matrix $\mathbf{e}$, whose components serve as the edge features, to the 7-dimensional vector $\mathbf{v} = (v_1, v_2, \ldots, v_7)^{\mathrm{T}}$. Excluding the last three components reduces the graph classifier to our classical benchmark, enabling a direct performance comparison. The aggregation step, represented by the function $\textsc{Agg}$, is chosen to be the permutation-invariant sum of all vertex feature vectors. This aggregation step constitutes the core of graph message passing, wherein messages are collected at each vertex from all other vertices in the graph, ensuring that the information from the entire jet substructure is incorporated into each constituent's representation.

For this particular graph classifier, only a single message-passing layer is necessary to achieve optimal performance. Following the aggregation step, we employ graph pooling, which reduces the entire graph to a single representative vector by combining information from all vertices. Specifically, we utilise the \textsc{Max} pooling operation after the graph messages have been aggregated, selecting the maximum value across each feature dimension. The final output from this graph architecture is a 7-dimensional vector, maintaining the same dimensionality as the input feature space. Crucially, the conventional non-linear transformation step that would typically be present in a standard GNN to convert the output into interpretable probabilities is not required here. This is because our entire framework represents a non-trainable approach, and the inference procedure utilising an appropriate distance metric will now be defined in the next section.

\subsection{Inference and Results: Tomography-enhanced Graph classifier}

Our input dataset consists of jets initiated either by light quarks or gluons (constituting the background class) or by the hadronic decay of top quarks ($t \to b q \bar{q}$), which serves as our signal class. We initially use a training dataset comprising 800,000 jets, equally divided between the two classes. These jets are processed through the untrained graph architecture to compute exactly two class means: a pair of 7-dimensional vectors that serve as the archetypal representations of the signal and background classes, respectively.

In the subsequent step, we utilise a separately held-out dataset consisting of 1 million jets, again equally divided between both classes, which serves as our test dataset for inference. Classification is then performed using the Mahalanobis distance as the discriminating metric, enabling us to assign each jet a class probability based on whether the output jet vector lies closer in the 7-dimensional hyperspace to the signal or background class archetype. The Mahalanobis distance is defined as:

\begin{equation}
d^2(\mathbf{x}) = (\mathbf{x} - \boldsymbol{\mu})^{\mathrm{T}} \boldsymbol{\Sigma}^{-1} (\mathbf{x} - \boldsymbol{\mu})
\label{eq:mahalanobis_distance}
\end{equation}

where $\mathbf{x}$ represents the 7-dimensional output vector from the graph classifier for a given jet, the terms $\boldsymbol{\mu}$ denote the mean vector for a particular class (either signal or background), and $\boldsymbol{\Sigma}^{-1}$ the inverse covariance matrix, both being computed from the training dataset.

For each output jet vector, we calculate the squared Mahalanobis distances to both the mean signal ($d_{\text{top}}^2$) and the mean background ($d_{\text{QCD}}^2$) vectors. An additional normalization factor $N = d_{\text{QCD}}^2 + d_{\text{top}}^2 + \epsilon$ is subsequently computed, where $\epsilon = 10^{-8}$ is chosen for numerical stability and to prevent divergences. Class probabilities are then assigned to each jet according to:

\begin{align}
P_\mathrm{QCD} &= \frac{d_\mathrm{QCD}^2}{N} \\
P_\mathrm{top} &= \frac{d_\mathrm{top}^2}{N}
\end{align}

This prescription ensures that class probabilities always sum to unity. Subsequently, classification becomes straightforward: jets lying closer to the mean top vector are classified as signal, and those closer to the background mean are classified accordingly. The proximity of a jet to its assigned class mean vector directly correlates with the confidence in that classification.

\begin{figure}[H]
\centering
\includegraphics[width=0.65\textwidth]{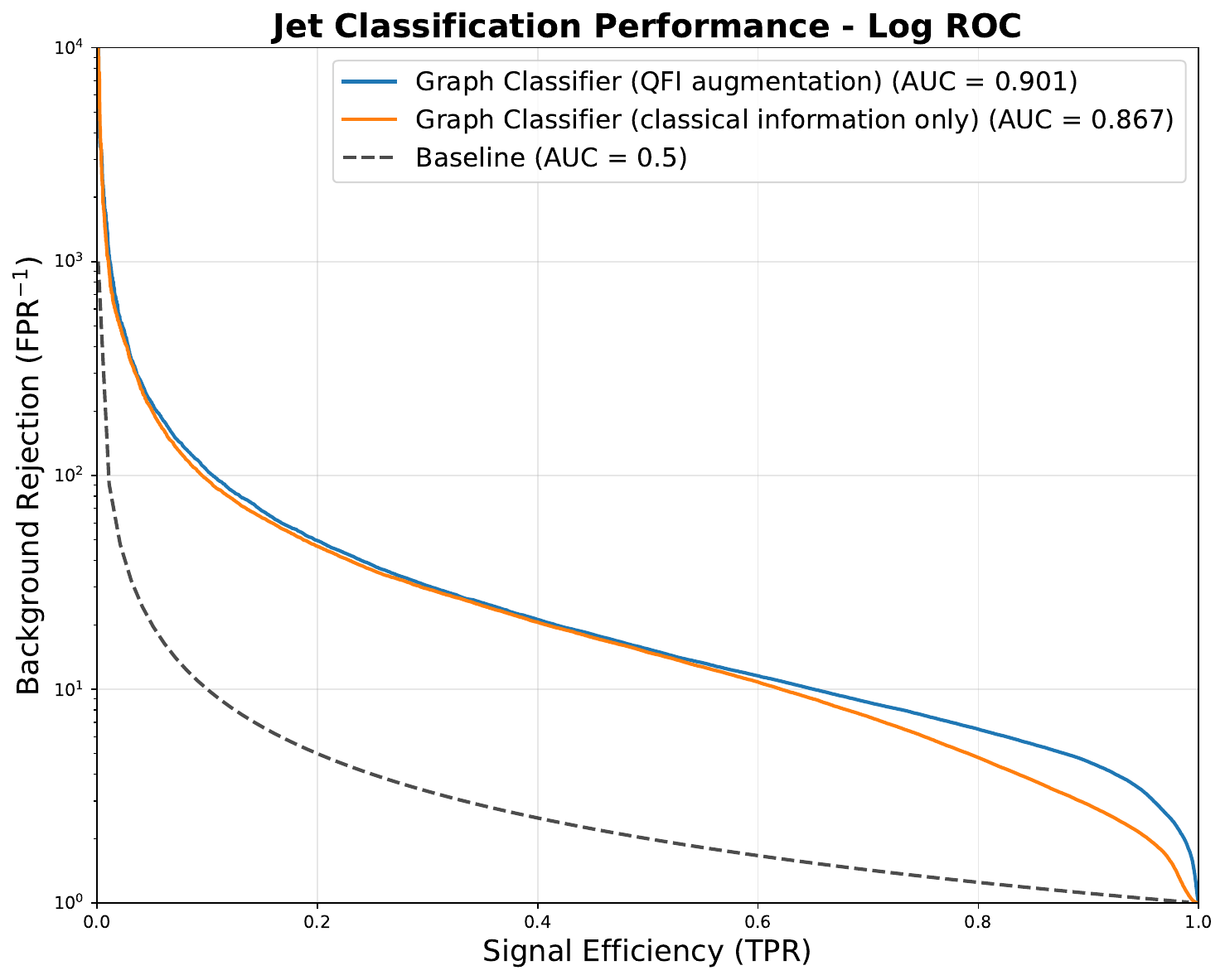}
\caption{AUC: classification performance of untrained graph classifiers with and without QFIM-based enhancement}
\label{fig:AUC_untrained_graphs}

\end{figure}

Finally, we construct the receiver operating characteristic (ROC) curve by calculating the true and false positive rates across various probability thresholds, and compute the area under this curve (AUC), which serves as a robust metric of classification performance. The comparison between the quantum-enhanced and non-enhanced versions of the graph classifier is presented in Figure~\ref{fig:AUC_untrained_graphs}, demonstrating that the AUC for the quantum-enhanced version clearly exceeds that of the classifier utilising purely kinematic information, thereby validating the efficacy of our proposed approach.

\section{Quantum-informed GNN}
\label{sec:qgnn}

The generalisation from an untrained graph classifier to a trained GNN is relatively straightforward and unlocks powerful capabilities that elevate classification performance to levels unattainable through simple mathematical transformations alone. The function $\psi^{(l)}$ of Eq.~\ref{eq:graph_message} is now replaced by a convolutional neural network~\cite{lecun1989backprop,schmidhuber2015deep} that accepts the same inputs as defined in Eq.~\ref{eq:input_feature_definition}. An additional optimisation we employ for more efficient utilisation of the input information is the creation of a compressed representation of the QFIM sub-matrix $\mathbf{e}$ present along each edge. This compression is achieved by passing the flattened $3 \times 3$ sub-matrix $\mathbf{e}$ through a trainable multi-layer perceptron (MLP)~\cite{rosenblatt1958perceptron} with three hidden layers $[16, 8, 4]$ and a final 3-dimensional output. This 3-dimensional embedding is subsequently used as input to $\psi^{(l)}$.

\begin{figure}[H]
\centering
\includegraphics[trim={0 18cm 0 0},clip,width=0.9\textwidth]{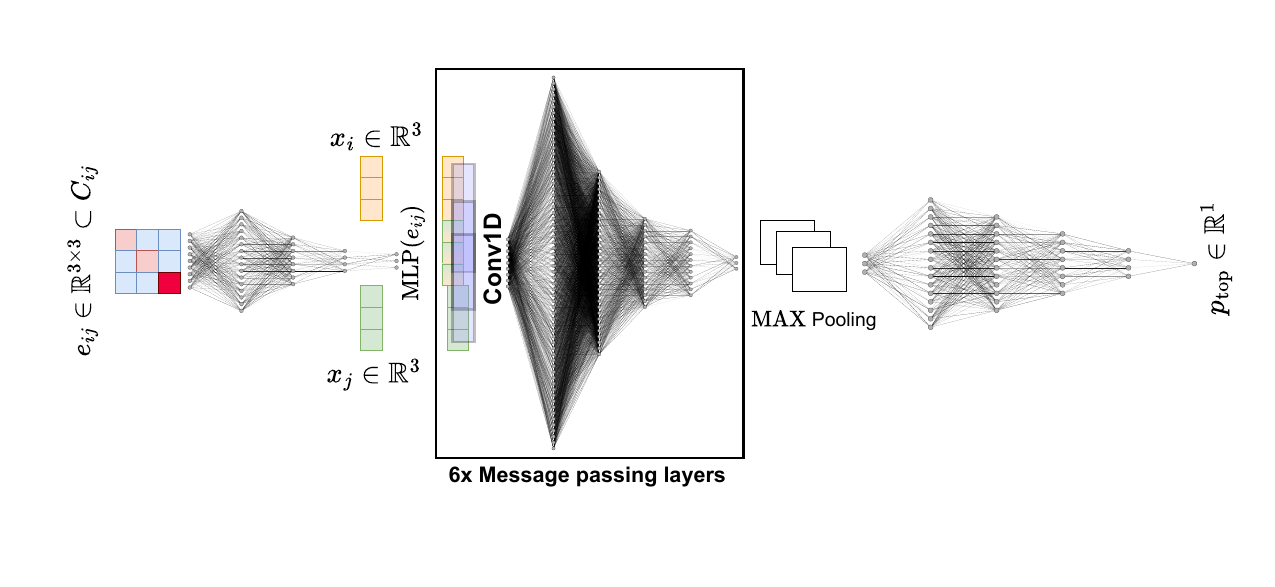}
\caption{Schematic of the quantum-enhanced GNN}
\label{fig:qGNN_schematic}
\end{figure}

The architecture of the convolutional neural network $\psi^{(l)}$ is optimized through a hyperparameter search and configured as follows: a \textsc{\textsc{Conv1D}} operation with four output channels and kernel size 5, followed by an MLP with four hidden layers $[64, 32, 16, 12]$ and a final output dimension of 3, matching the input dimensionality. The preservation of dimensionality between input and output enables the implementation of residual connections within the graph structure. Residual connections facilitate gradient flow during training and allow the network to learn identity mappings when beneficial, thereby enhancing both training stability and model expressivity. The complete message function is defined as:

\begin{equation}
\psi^{(l)} = \textsc{MLP}\left(\textsc{\textsc{Conv1D}}\left(h_i, h_j, \Vec{\boldsymbol{\lambda}} \cdot \textsc{MLP}(e_{ij})\right)\right)
\label{eq:trained_message_function}
\end{equation}

where $\Vec{\boldsymbol{\lambda}}$ is a learnable temperature vector that allows the QFIM to influence the message passing operation. The network employs a total of 6 such message passing layers to capture complex multi-scale interactions that define the jet substructure. A schematic of the complete quantum information-enhanced GNN is shown in Figure \ref{fig:qGNN_schematic}.

The classical benchmark is constructed by replacing $\mathbf{e}$ with a null matrix, effectively removing all quantum information enhancement while maintaining identical network architecture. Following the message computation using the 1D convolutional network and embedding MLP, the graph aggregation and pooling steps remain identical to those employed in the untrained classifier: namely, a \textsc{Sum} aggregation followed by \textsc{Max} pooling. Finally, an additional MLP, with four hidden layers $[16, 12, 8, 4]$ followed by a softmax output layer, converts the aggregated and pooled representation into a directly interpretable probability score, following standard GNN practice. The network in both cases is trained using the Binary Cross Entropy (BCE) loss and the \textsc{Adam} optimiser with a learning rate initially set to $0.005$ and allowed to decay gradually, based on the improvement shown on the validation dataset. The training dataset comprises 1 million jets, whereas the validation dataset is of size 200,000 jets. The test dataset contains 1.2 million jets. In all three cases, the datasets are equally divided between the background (light quark- or gluon-initiated jets) and signal (\ttbar decay jets) classes. 

\begin{figure}[H]
\centering
\includegraphics[width=0.65\textwidth]{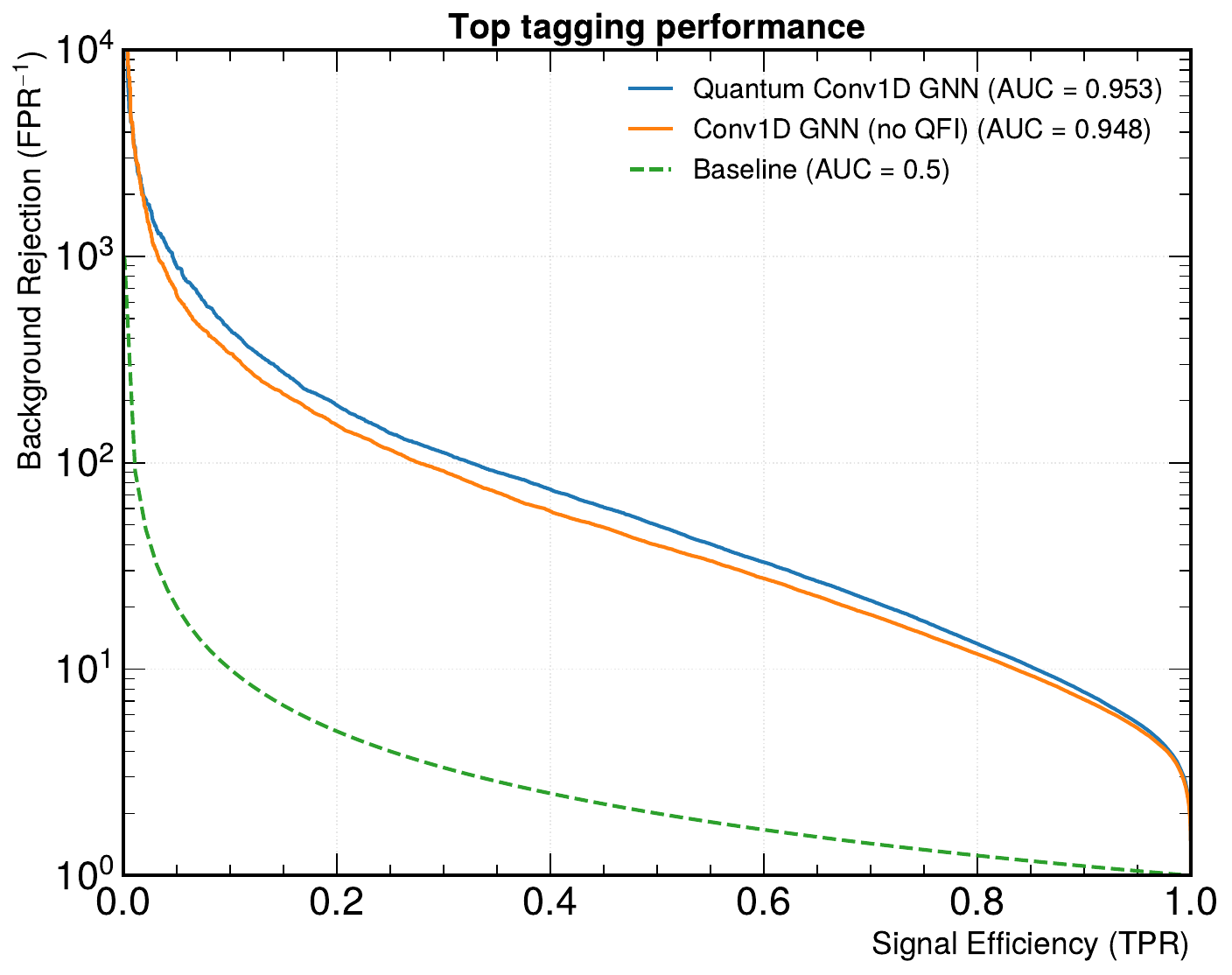}
\caption{AUC: classification performance of trained GNNs with and without QFIM-based enhancement}
\label{fig:AUC_GNN}
\end{figure}

\begin{figure}[H]
\centering
\includegraphics[width=0.95\textwidth]{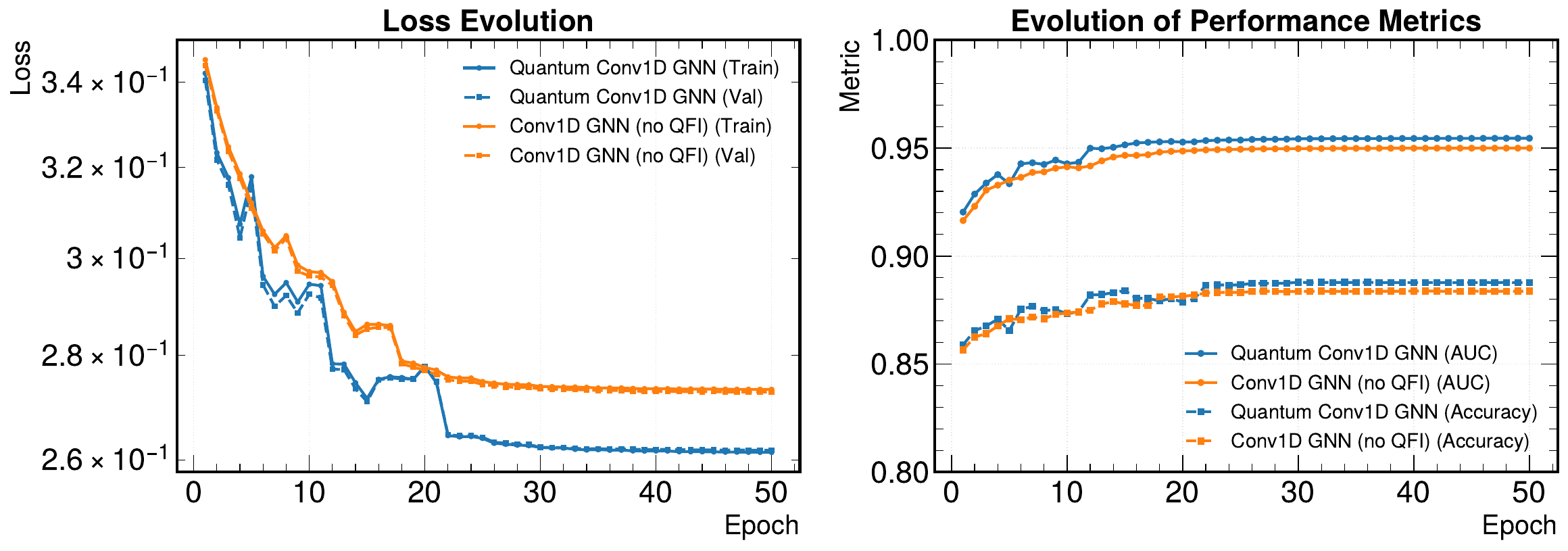}
\caption{Evolution of metrics and loss function for both GNNs}
\label{fig:loss_GNN}
\end{figure}

\section{Summary and Conclusions}
\label{sec:summary}

Quantum-informed neural networks provide an interpretable and practically useful route to jet tomography. A one-particle-one-qubit embedding, paired with the Quantum Fisher Information Matrix, exposes physically meaningful correlations and propagates them into both simple baselines and scalable graph learning. Following the structure outlined in Sec.~\ref{sec:intro}, the analysis proceeded in four steps with concrete findings at each stage. Collectively, our approach establishes a pipeline from quantum geometry to high-performing classifiers for collider data.

First, we focus on the tomographic premise. Encoding each of the leading constituents on its own qubit and computing the full, non-block-diagonal QFIM reveals class-dependent geometry in state space. Ensembles of QCD jets are characterised by QFIMs dominated by near-diagonal structure, consistent with largely quasi-independent splittings. In contrast, hadronic top jets exhibit pronounced off-diagonal elements that reflect three-pronged, colour-connected dynamics. Recovering these inter-dependencies required the complete QFIM rather than block diagonal approximations. This shows that the QFIM is a basis-independent, interpretable observable that magnifies the correlation patterns expected from the underlying colour flow.

Second, we tested whether this quantum geometric summary can separate classes without training. A prototype classifier that vectorises the per-event QFIM and scores jets by distance to class prototypes already achieves strong separation with an AUC of 0.883. Seeding a shallow network with the class mean QFIMs preserves the performance gain and, more importantly, stabilises and accelerates optimisation relative to random initialisation, with both initialisations converging to nearly identical, QFIM-like prototypes and reaching AUCs around 0.917. These observations show that the QFIM contain class information and also acts as an informative prior that improves optimisation dynamics by aligning the model with the intrinsic geometry of the data.

Third, we elevated the tomographic object into a fixed, non-trainable graph representation. Treating constituents as nodes and inserting QFIM sub-matrices as edge features creates an informed adjacency that guides message passing. With node features built from simple relative kinematics, event-level inference via a Mahalanobis distance rule improves the AUC from 0.867 for a kinematics-only baseline to 0.901 when QFIM features are included. The construction uses up to ten constituents per jet and evaluates on held-out samples at the million-event scale, so the observed gain can be traced to the quantum geometric correlations themselves rather than to learned parameters.

Finally, we generalised to a trained, quantum-informed GNN. The message function combines a \textsc{Conv1D} operation on node pairs with a compact, learned embedding of the 3×3 QFIM block attached to each edge, with residual connections used to maintain stable updates. Relative to an otherwise identical architecture with the QFIM input nullified, the quantum informed network improves top tagging AUC from 0.948 to 0.953 and converges faster and more smoothly in both loss and validation metrics. These results indicate that quantum encoded correlations help both accuracy and optimisation behaviour, while integrating naturally with standard graph learning practice.

Thus, the 1P1Q embedding and the QFIM reveal class-dependent correlation geometry, the untrained prototype confirms that the geometry alone separates classes and shapes the optimisation landscape, and the fixed graph demonstrates that QFIM augmentation sharpens classical summaries in a parameter-free setting. The trained GNN shows that quantum-informed edges translate into measurable performance gains and improved training dynamics at scale. In all cases, the quantum algorithms are compact and shallow, which preserves compatibility with near-term devices and keeps the approach attractive for hybrid classical quantum setups.

Beyond metrics, our approach shows that the QFIM provides an interpretable observable and a design prior for neural networks. As a quantum metric, the QFIM connects directly to sensitivity and natural gradient notions, so it offers a principled way to encode physics-aligned inductive bias into classical architectures. The empirical stability and faster convergence we observe are consistent with this view and suggest that quantum geometry can regularise optimisation while maintaining a clear physical interpretation of what the network learns.

We find that QINNs for jet tomography expose and exploit correlation structures in a way that is both interpretable and scalable. They improve classical baselines in fixed and trainable settings, and they remain compatible with near-term hardware. This combination of interpretability, performance, and practicality approaches a credible candidate for use in collider data analysis.

\paragraph{Acknowledgement:} The authors acknowledge the support of Schmidt Sciences, the Alexander von Humboldt Foundation, and IPPP Durham, as well as the usage of computing resources on the \texttt{deepthought} GPU cluster at the Institute of Experimental Particle Physics (ETP), KIT, and the RCS Cluster at Imperial College London.

\bibliographystyle{inspire}
\bibliography{apssamp.bib}% Produces the bibliography via BibTeX.
\end{document}